\documentclass[conference]{IEEEtran}
\usepackage{amsmath,amssymb,amsfonts}
\usepackage{algorithmic}
\usepackage{graphicx}
\usepackage{textcomp}
\usepackage{xcolor}
\usepackage{placeins}
\usepackage[hidelinks]{hyperref}
\def\BibTeX{{\rm B\kern-.05em{\sc i\kern-.025em b}\kern-.08em
    T\kern-.1667em\lower.7ex\hbox{E}\kern-.125emX}}
\usepackage{comment}
\usepackage{graphicx}
\usepackage{booktabs}
\usepackage{enumitem,amssymb}
\usepackage{multirow}
\usepackage{colortbl}
\newlist{todolist}{itemize}{2}
\setlist[todolist]{label=$\square$}
\usepackage{pifont}
\usepackage{caption}
\usepackage{subcaption}
\usepackage{pgfplots}
\usepackage{bbm}
\usepackage{acronym}

\newcommand{\method}{\textsc{SALAD}}
\newacro{LM}{Language model}
\newcommand{\newpara}[1]{\vspace{0.48em}\noindent\textbf{#1}~}
\newacro{E2E}{end-to-end}
\newacro{AR}{autoregressive}
\newcommand{\AR}{\ac{AR} }
\newacro{NAR}{non-autoregressive}

\newacro{TTS}{text-to-speech}
\newacro{T2A}{text-to-acoustic}
\newacro{T2S}{text-to-semantic}
\newacro{S2A}{semantic-to-acoustic}
\newacro{GMM}{Gaussian Mixture Model}
\usepackage[backend=biber,style=ieee]{biblatex}
\begin{document}

\title{Speech Synthesis From Continuous Features Using \\ Per-Token Latent Diffusion}

\author{
\IEEEauthorblockN{
Arnon Turetzky$^{1,2,\S}$,
Avihu Dekel$^{2,\S}$,
Nimrod Shabtay$^{2,3}$,
Slava Shechtman$^{2}$,\\
David Haws$^{2}$,
Hagai Aronowitz$^{2}$,
Ron Hoory$^{2}$,
Yossi Adi$^{1}$
}
\IEEEauthorblockA{$^{1}$The Hebrew University of Jerusalem}
\IEEEauthorblockA{$^{2}$IBM Research}
\IEEEauthorblockA{$^{3}$Tel Aviv University}
\IEEEauthorblockA{$^{\S}$Core Contributors}
}

\maketitle
\begin{abstract}
We present \method, a zero-shot \ac{TTS} autoregressive model operating over continuous speech representations. \method~utilizes a per-token diffusion process to refine and predict continuous representations for the next time step. We compare our approach against a discrete variant of \method~as well as publicly available zero-shot TTS systems, and conduct a comprehensive analysis of discrete versus continuous modeling techniques. Our results show that \method~achieves superior intelligibility while matching the speech quality and speaker similarity of ground-truth audio.
\end{abstract}

\begin{figure*}[t!]
    \centering
    \includegraphics[width=0.98\linewidth]{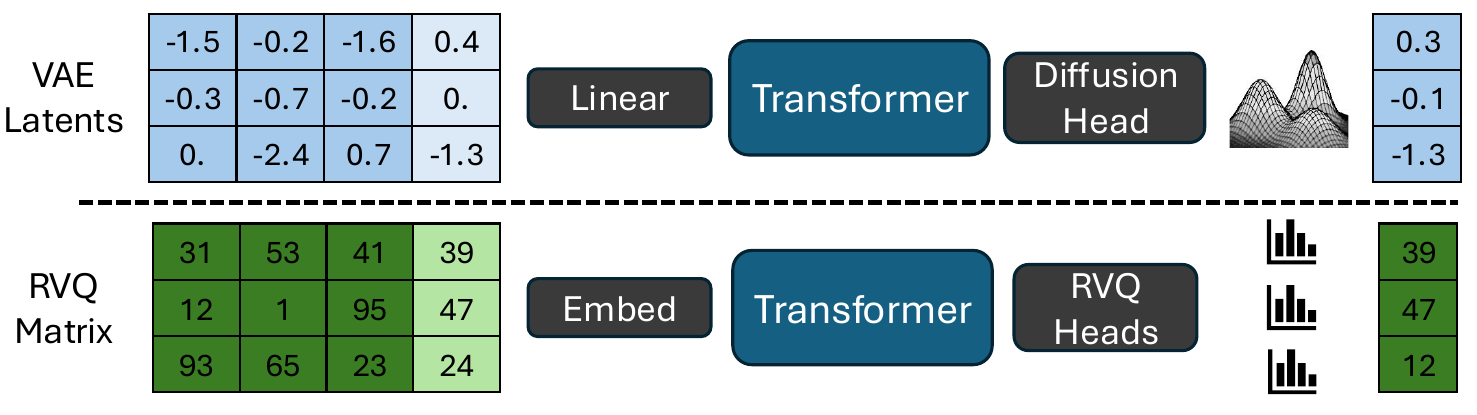}
    \caption{Continuous vs. discrete modeling}
    \label{fig:discrete_vs_cont}
    \vspace{-0.2cm}
\end{figure*}
\section{Introduction}
\label{sec:intro}

Discrete \AR models have become the leading approach in natural language processing. Their remarkable success has motivated researchers to extend these models to image and speech domains. Modeling continuous domains with discrete approaches typically involves quantization. In image generation, quantization is often realized using discrete autoencoders \cite{VQVAE}, which are further optimized with adversarial losses \cite{VQGAN,DALLE,Make-a-scene}. In audio generation, techniques like \textit{Residual Vector Quantization} (RVQ) \cite{Soundstream,Encodec} iteratively quantize the residual, producing multiple discrete codes per frame that can be modeled autoregressively \cite{VQGAN,VALLE,Musicgen}. 

While presenting impressive results, discrete approaches have several limitations. First, modeling multiple dependent streams simultaneously is inherently complex and often requires additional strategies like two-stage modeling \cite{VALLE,VallE2,VallEX} or token interleaving \cite{Musicgen}. Second, similarity among consecutive codec codes has been shown to cause robustness issues, leading to continuous stretches of silence or persistent noise~\cite{VallE2}. Third, quantization methods inherently impose a fidelity \textit{upper-bound}, constrained by continuous representations such as mel-spectrogram features~\cite{puvvada2024discrete}. This raises the question: \emph{Can we optimize an \AR model to directly generate high-quality speech over continuous features?}

Recently, several studies have explored language modeling with continuous-valued tokens. In image generation, GIVT~\cite{GIVT} proposes representing the continuous distribution of images obtained by a VAE using a \ac{GMM}, while AR-Diffusion~\cite{ARDiffusion} suggests a per-token image diffusion head to model similar representations. For speech generation, \cite{MELLE} presented an \AR modeling approach over mel spectrograms, incorporating a Gaussian sampling module followed by a post-net. Concurrently with our work, \cite{lin2025continuous} propose a similar modeling approach to GIVT but for spoken data.

Inspired by AR-Diffusion~\cite{ARDiffusion}, we propose \method~(Speech synthesis with Autoregressive LAtent Diffusion), an autoregressive per-token latent diffusion model for zero-shot speech synthesis over continuous features. By operating directly in continuous space, \method~eliminates the need for signal quantization, thus enabling higher-fidelity speech generation. As an autoregressive model, \method~also naturally supports \textit{variable-length} outputs, addressing a fundamental challenge not encountered in fixed-length image generation methods. We use semantic tokens~\cite{Hubert,w2v,w2vbert} which are not used for signal synthesis but rather to enhance robustness and define the generation-stopping condition.

We evaluate \method~against the \textit{discrete} representation alternative employing RVQ quantization, specifically following the delay prediction pattern introduced by~\cite{Musicgen} (see Figure~\ref{fig:discrete_vs_cont}). Additionally, we compare \method~with two leading publicly available zero-shot TTS systems, XTTS~\cite{XTTS} and VoiceCraft~\cite{peng2024voicecraft}. Our experimental results demonstrate that \method~achieves superior intelligibility and maintains speech quality and speaker similarity comparable to ground-truth audio, as confirmed by both objective metrics and subjective listening tests
\footnote{Samples available at: \url{https://s3.us-south.objectstorage.softlayer.net/zk-wav-data/Webpages/SynthesisPerTokenLatentDiffusion/index.html}}.

\newpara{Our contributions:} 
(i) We introduce \method, the first autoregressive per-token latent diffusion model for zero-shot speech synthesis directly over continuous acoustic representations, eliminating the need for discrete quantization. 
(ii) We empirically demonstrate that continuous latent diffusion modeling can surpass discrete modeling approaches in intelligibility and match them in quality and speaker similarity. 
(iii) Through comprehensive experiments and ablation studies, we analyze the advantages and trade-offs of continuous versus discrete speech modeling, providing practical insights into the design of zero-shot speech generation models.

\section{Related Work}
\label{sec:related_work}
\newpara{Zero-Shot TTS.} Inspired by the success of in-context learning, there has been significant interest in zero-shot \ac{TTS} systems that generalize to unseen speakers during inference, offering flexibility and improved quality~\cite{VALLE}. Zero-shot TTS methods typically formulate the task as language modeling over text and audio tokens, leveraging short speaker prompts to synthesize speech aligned with the target speaker’s voice characteristics~\cite{Voicebox,NaturalSpeech2,BaseTTS,peng2024voicecraft}. Our method follows this paradigm but operates directly over continuous speech representations.

\newpara{Semantic Tokens.}
Quantized embeddings from self-supervised audio models, termed semantic tokens~\cite{Hubert,w2v,w2vbert}, capture phonetic and prosodic information beneficial for speech synthesis~\cite{SPEAR,MakeAVoice,Soundstorm}, unconditional audio generation~\cite{AudioLM}, and multimodal text-audio tasks~\cite{AudioPalm}. \method~utilizes semantic tokens as an auxiliary representation defining a precise stopping condition for continuous acoustic modeling.

\newpara{RVQ Codes Prediction.}
Discrete audio generation methods commonly utilize \textit{Residual Vector Quantization} (RVQ), involving multiple quantization layers~\cite{Soundstream,Encodec}. Methods such as AudioLM~\cite{AudioLM} flatten RVQ codes into sequences, whereas others, like Vall-E~\cite{VALLE}, employ two-stage prediction approaches or token interleaving strategies~\cite{Musicgen}. Differently, \method~ directly predicts a continuous latent space, thus avoiding the need to predict multiple residuals codes. 

\newpara{Continuous Models.}
When learning a continuous distribution, recent works typically use diffusion models, which were developed to sample from complex continuous probability distributions, inspired by non-equilibrium thermodynamics \cite{DDPM}.
Several works attempt to synthesize speech using a diffusion process, which has the challenge of generating variable length outputs~\cite{Diffwave,Wavegrad,GradTTS}. For that end, most diffusion-based works rely on a duration predictor that predicts the audio length in advance, which might be inferior to determining the length on-the-fly during synthesis~\cite{NaturalSpeech2,Voicebox}.
MELLE~\cite{MELLE} predicts Mel spectrograms autoregressively using a Gaussian sampling module, and parameterizes the next frame using a Gaussian distribution, which restricts it to learn only unimodal distributions.
MELLE relies on an additional binary classifier that indicates when to stop, which is a highly imbalanced classification problem.
In contrast, \method\ operates on VAE latent tokens, which allows sampling diverse inputs while training, and uses a diffusion head, capable of modeling multimodal distributions. \method\ relies on semantic-tokens to determine the stopping condition, a more balanced representation which also provides contextual information.

\begin{figure*}[t!]
\centering
    \begin{subfigure}[b]{0.38\textwidth}
    \includegraphics[width=\linewidth]{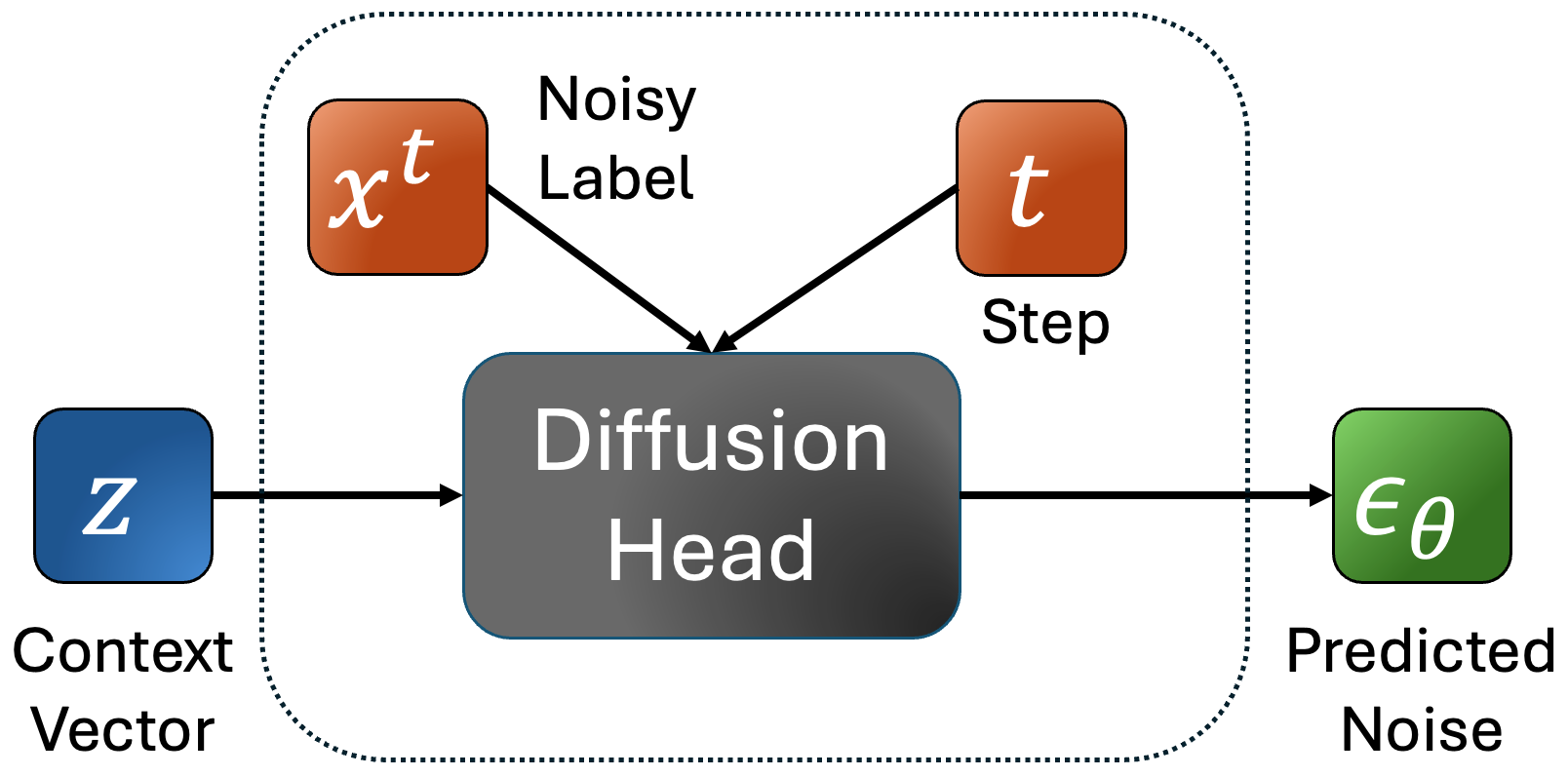}
    \caption{Training}
    \label{fig:diffusion_head_train}    
    \end{subfigure}
    \includegraphics[width=0.012\linewidth,height=0.25\linewidth]{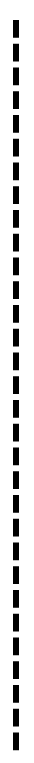}
    \begin{subfigure}[b]{0.38\textwidth}
        \includegraphics[width=\textwidth]{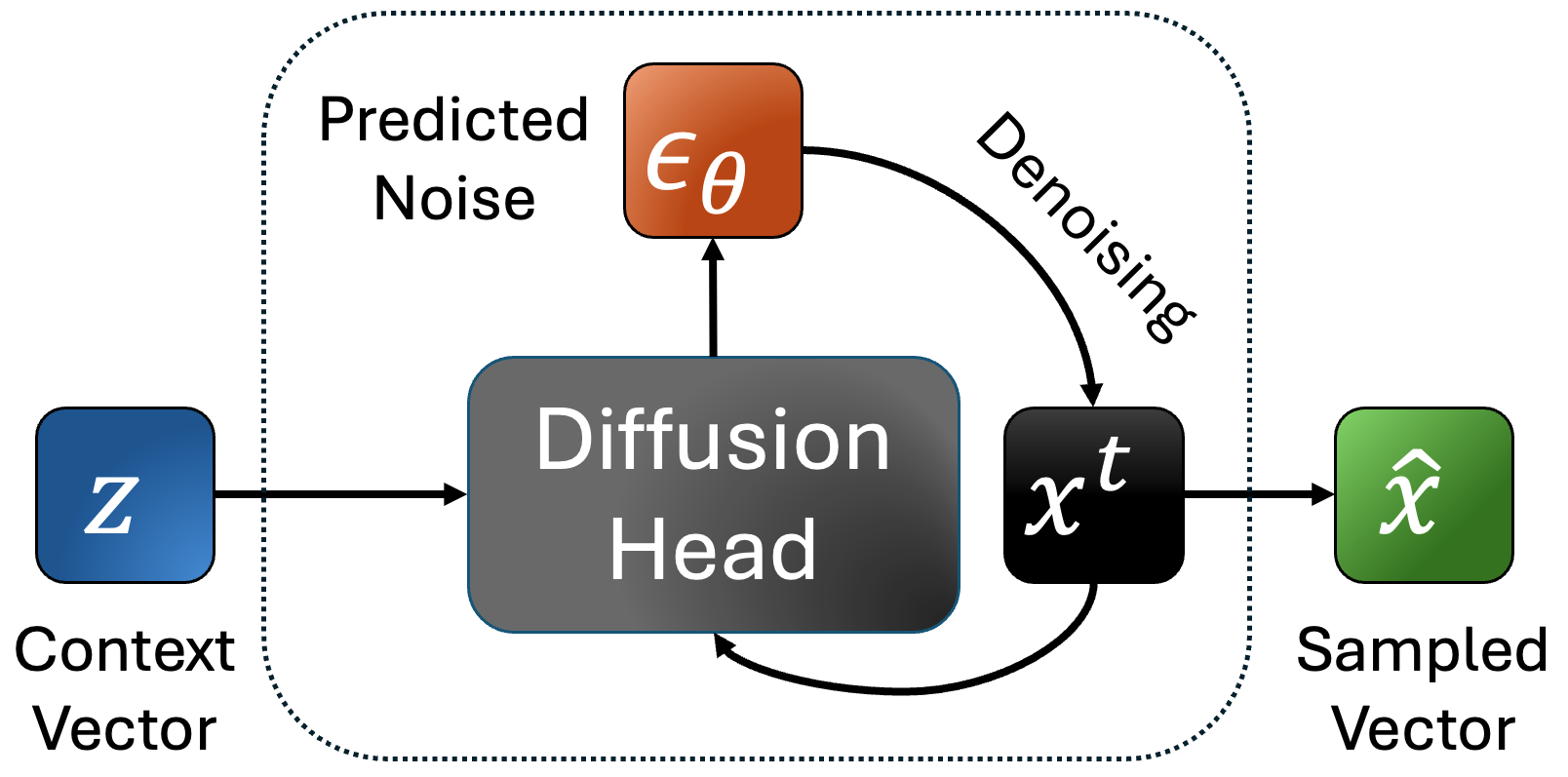}
        \caption{Inference}
        \label{fig:diffusion_head_infer}    
    \end{subfigure}   
    \caption{The per-token diffusion head}
    \label{fig:diffusion_head}    
    \vspace{-0.2cm}
\end{figure*}

\section{Background}
\label{sec:background}

\newpara{Definitions.}
We denote the raw audio sequence as $\boldsymbol{a}=(a_1,...,a_m)$ where $a_i\in [-1,1]$ with sampling rate $f_S$.
The text is $\boldsymbol{y}=(y_1,..,y_k)$ where $y_i\in\mathcal{A}$, and $\mathcal{A}$  is the text vocabulary.
We obtain compressed audio representations using a variational autoencoder (VAE), trained with adversarial losses to obtain high-fidelity reconstructions. 
The VAE's encoder $\mathcal{E}$ predicts a sequence of means and variances of normal distribution:
$(\mu_1,...,\mu_n), (\sigma^2_1,...,\sigma^2_n) = \mathcal{E}(\boldsymbol{a})$
where $\sigma_i,\mu_i\in \mathbb{R}^d$ and $d$ is the VAE bottleneck dimension. 
The VAE downsamples the sequence with a stride $r$.
We sample $x_i \sim \mathcal{N}(\mu_i,\sigma_i^2)$ and 
denote $\boldsymbol{x}=(x_1,..,x_n)$ as the continuous \textit{acoustic tokens}. Hereafter, we refer to these continuous VAE latents interchangeably as acoustic tokens or continuous acoustic representations, following prior usage in speech-diffusion literature\cite{MELLE}. The VAE's decoder $\mathcal{D}$ is used for reconstruction 
$\hat{a}_1,...,\hat{a}_m = \mathcal{D}(x_1,..,x_n).$
We also extract semantic tokens
and denote them by $\boldsymbol{w}=(w_1,..,w_m)$, which have the same downsampling stride as the VAE.
Our goal is to predict the audio based on the desired text and the speaker prompt. Denoting the speaker prompt latent features as $\boldsymbol{s}=s_1,...,s_p$, our training objective can be formulated by: $p(\boldsymbol{x} | \boldsymbol{y}, \boldsymbol{s})$.

\newpara{Diffusion Process.}
\label{subsec:diffusion}
A diffusion process starts from a continuous signal, and gradually destroys it using a forward noise process. 
Our method performs latent diffusion, and attempts to predict the VAE latent vectors $x_1,...,x_n$. 
Given noising coefficients $\beta_0,...,\beta_T$ and some continuous vector $x$, we define $x^0=x$ and $\epsilon\sim\mathcal{N}(0,I)$; the Markov structure is
$x^t=\sqrt{1-\beta_t}x^{t-1}+\sqrt{\beta_t}\epsilon$.
This iterative denoising process can be simplified. By defining $\alpha_t=1-\beta_t$ and $\bar{\alpha}_t=\prod^t_{i=1}\alpha_i$, we get that
$x^t=\sqrt{\bar{\alpha_t}}x+\sqrt{1-\bar{\alpha_t}}\epsilon$.
The diffusion process is often defined such that $\bar{\alpha}_T \to 0$ and $x^T$ distributes closely to the standard normal distribution. 
Diffusion models $\epsilon_\theta$ are trained to perform the reverse diffusion process, which denoises the corrupted signal by predicting the added noise. Their denoising loss is defined as
$\mathcal{L} (x) = \mathbb{E}_{\epsilon,t}\left[ \| \epsilon - \epsilon_{\theta}(t,x^t)\|^2\right]$.
Most diffusion models operate on a sequence $x_1,..,x_n$ and attempt to denoise all tokens in parallel using $\epsilon_\theta (t, x_1^t,...,x_n^t)$.

\newpara{Per-Token Diffusion Head.}
\label{subsec:diffusion_head}
\cite{ARDiffusion} proposed an MLP diffusion head for image generation. 
Unlike standard diffusion models, the diffusion head denoises each token \textit{independently}, which gives additional flexibility when defining the conditioning information (e.g., predicting on previously predicted tokens).
The authors rely on a transformer model $\Theta$ that extracts contextual per-token conditioning vectors  $z_1,..,z_n$ based on the input features and optional context vectors that we denote by $C$ 
\begingroup
\setlength{\abovedisplayskip}{3pt}
\setlength{\belowdisplayskip}{3pt}
\[
\boldsymbol{z}=z_1,...,z_n = \Theta(C, x_1,...,x_n).
\]
\endgroup

The diffusion head (noise estimator) $\epsilon_{\theta}$ takes a contextual conditioning vector $z$ and attempts to model the continuous distribution $p(x|z)$. 
Given a target token $x$, a diffusion process is being applied conditioned on $z$. The loss is:
\begin{equation}\label{eq:diffusion_loss}
\mathcal{L} (x,z) = \mathbb{E}_{\epsilon,t}\left[ \| \epsilon - \epsilon_{\theta}(x^t,t,z)\|^2\right].
\end{equation}
During training, $t\sim[T],\epsilon\sim\mathcal{N}(0,I)$ is sampled for each token $x$, resulting in noisy targets $x^t$, and $\mathcal{L} (x,z)$ is minimized. See Figure~\ref{fig:diffusion_head_train}.
The denoising network is trained jointly with the transformer $\Theta$, and the gradient with respect to $z$ is propagated to the transformer.
$K$ different values of $t,\epsilon$ may be sampled for a given context vector and target $z,x$, with the additional complexity of just the MLP head rather than the entire model.
During inference, a continuous vector is sampled from $x^T\sim\mathcal{N}(0,I)$ and the final result is obtained from the reverse diffusion process (see Figure~\ref{fig:diffusion_head_infer}):
\begin{equation}\label{eq:diffusion_inference}
x^{t-1}= \frac{1}{\sqrt{\alpha_t}} \left(x^t - \frac{\beta_t}{\sqrt{1-\bar{\alpha}_t}}\epsilon_\theta(x^t,t,z)\right)+\sqrt{\beta_t}\epsilon.
\end{equation}
\begin{figure*}
    \centering
\includegraphics[height=0.16\linewidth]{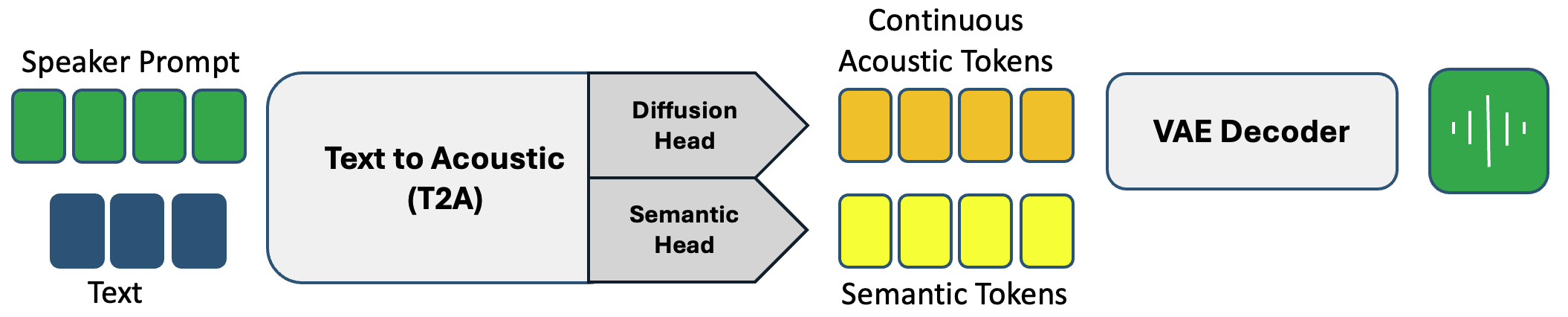}
    \caption{System overview. Conditioned on text $t$ and a speaker prompt $s$, \method{} emits semantic tokens $w$ and acoustic latents $x$, which a VAE decoder converts to waveform.}
    \label{fig:text_to_acoustic}
\end{figure*}

\section{\method}
\label{sec:salad}
\label{sec:method}
\method~is an end-to-end zero-shot \ac{TTS} model, that given text and a speaker prompt autoregressively predicting continuous VAE latents $\boldsymbol{x}$ using per-token latent diffusion.  Our method utilizes semantic tokens $\boldsymbol{w}$ as an auxiliary representation that provides contextual information and determines the stopping condition. 
We project both the semantic and acoustic representations into a shared space, and sum them to obtain a fused representation. We adopt the delay pattern suggested by \cite{Musicgen}, such that each acoustic token $x_i$ is predicted based on the semantic token $w_i$, for which we define $r_i=f(w_i, x_{i-1})$. 

Formally, model predicts both discrete semantic tokens and continuous acoustic tokens autoregressively, based on the text and speaker, using a causal transformer:
\begingroup
\setlength{\abovedisplayskip}{3pt}
\setlength{\belowdisplayskip}{3pt}
\begin{multline}
\label{eq:ar}
p\!\big((w_1,x_0),\ldots,(w_n,x_{n-1}) \mid t,s\big) \\
= \prod_{i=1}^{n} p\!\big((w_i,x_{i-1}) \mid t,s,r_1,\ldots,r_{i-1}\big).
\end{multline}
\endgroup

For each autoregressive step $i$, we extract contextual features from our transformer backbone, based on the text and speaker prompt $z_i = \Theta(t,s,r_1,...r_i)$, which is used to predict $w_{i+1}$ using the cross-entropy loss $L_s$ by the MLP head, and $x_{i}$ using the diffusion loss $L_a$ by the diffusion head (see the system overview in Figure~\ref{fig:text_to_acoustic}). We halt the generation once the semantic prediction head samples an EOS token. 
We note that audio duration is predicted on the-fly based on the model's predictions, unlike most diffusion-based TTS models, where the audio duration is predetermined.

\section{Experiments}
\label{sec:exp_setup}

\subsection{Experimental Setup}
\newpara{Datasets.}
We train our models on the English subset of multi-lingual LibriSpeech (MLS) \cite{pratap2020mls}, which contains 10M examples of 10-20 seconds, resulting in 45K hours. 
To avoid over-exposure of a few speakers, we limit the maximal number of utterances per speaker to 10K, resulting in 5.2M examples.
We evaluate all models on LibriSpeech \textit{test-clean} \cite{panayotov2015librispeech}, which consists of 2620 utterances by 40 speakers. 
All speakers in the test set are excluded from the training set.
We filter the dataset to utterances with lengths of 8-25 seconds, and then limit to at most 15 samples per speaker, resulting in 564 utterances for evaluation. 

\newpara{Tokenization.} 
To derive acoustic tokens, we train continuous $\beta$-VAE-GAN, with a varying bottleneck dimension $d\in \left\{ 8,16,24,32 \right\}$, and set the KL-divergence regularization to $\beta=5\cdot10^{-5}$, as done in \cite{GIVT}. 
We also train discrete RVQ-GAN models with $q\in \left\{4,8,12\right\}$ codebooks, each with 1024 entries. In addition, we apply quantizer dropout~\cite{Soundstream} with $p=0.5$. 
All compression models are trained on MLS-English, DAPS, LibriTTS, LibriTTS-R and LJ-Speech, which balance between high and mid quality recordings~\cite{shechtman24_interspeech}. 
The all-training hyperparameters follow the original recipe proposed by \cite{DAC}.
We extract semantic tokens by quantizing the embeddings of the 11th layer of W2V-BERT \cite{barrault2023seamless} using minibatch K-means with 1024 centroids. 

\newpara{Architecture.} We use a transformer backbone with $d=1024$, $d_{ff}=4096$, $24$ layers, $16$ heads, sinusoidal positional embedding, GeLU activation, and a dropout rate of $0.1$, resulting in models with roughly $350$M parameters. 
VAE embeddings are projected using a linear layer, while RVQ tokens are embedded using $Q$ lookup tables, which are summed into a single embedding. 
We use Classifier-Free Guidance (CFG)~\cite{CFG} and randomly omit the speaker prompt with $p=0.1$ during training.

RVQ codes are predicted using a $Q$ MLP heads with four hidden layers. For discrete distributions we apply top $k=10$ sampling, with a temperature of $\tau=1$, a repetition penalty of $1.05$, and a CFG scale of $\alpha=3$.

We use a diffusion process with $T=1000$ steps, where betas are logarithmicly spaced between $\beta_{0}=2e-4$ and $\beta_{T}=0.03$.
Our per-token diffusion head is an MLP network with 12 residual layers, that predicts the noise $\epsilon$ given the transformer embedding vector $z$, the noisy input $x^t$, and the diffusion step $t$.
Each residual block consists of layer normalization, linear layer, SiLU activation, and dropout with $p=0.1$. During inference, we apply $20$ diffusion steps for sampling, with a default noise scale of $1$. 
We use the AdamW optimizer, with $lr_{max}=3e-4$ and $lr_{min}=3e-5$, weight decay $0.1$, and a clip gradient norm of $1$, and train with FP16 mixed precision. 
We linearly warm up the learning rate from $lr_{min}$ across 32K iterations to $lr_{max}$ and decay the learning rate back to $lr_{min}$ over 300K steps using a cosine schedule. 
Each global batch size has $\sim$150K acoustic tokens. Each model was trained with 8 A100 80GB GPUs.

\begin{figure*}[t]
\centering
\begin{subfigure}[b]{0.45\textwidth}
    \includegraphics[width=\textwidth]{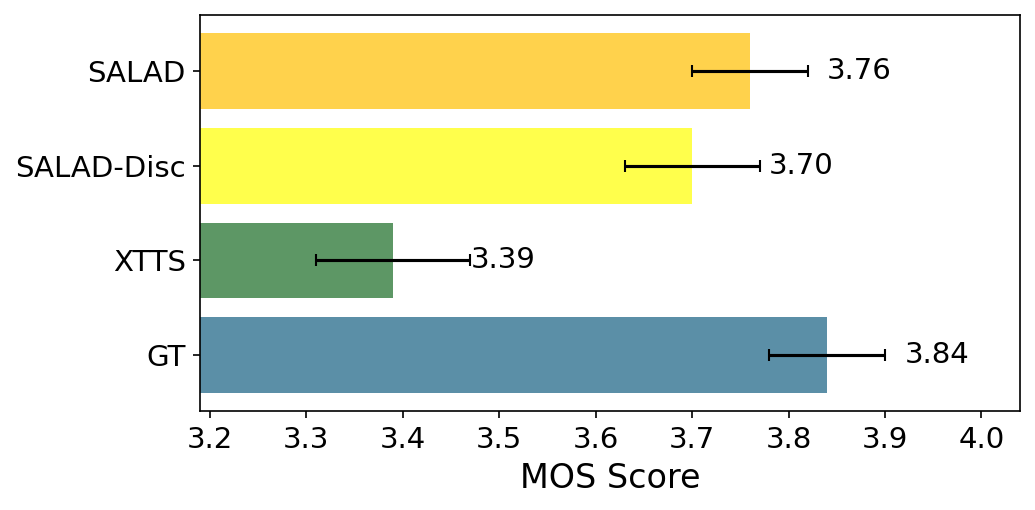}
    \caption{MOS Listening Test (1-5 scale)}
    \label{fig:mos_test}
\end{subfigure}
\hspace{0.03\textwidth}
\begin{subfigure}[b]{0.45\textwidth}
\includegraphics[width=\textwidth]{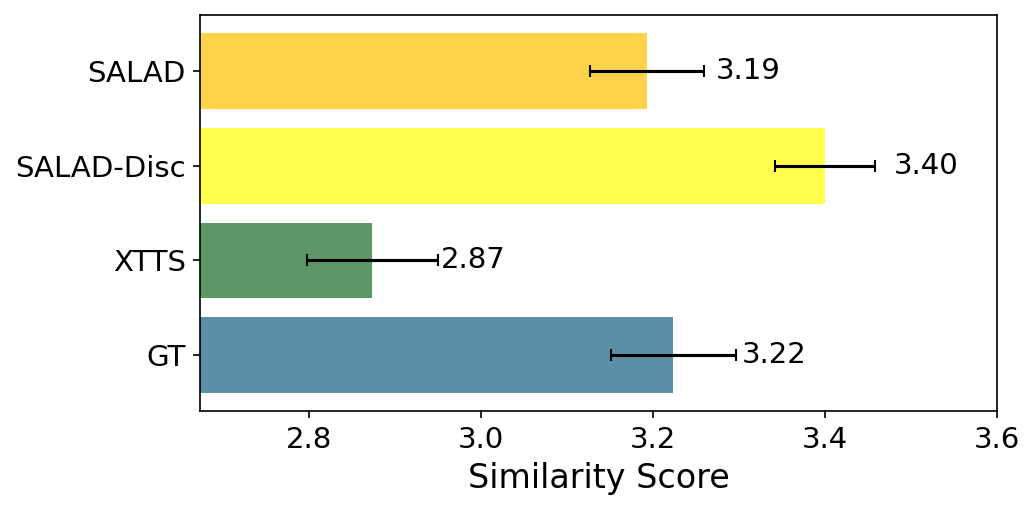}
\caption{Speaker Similarity Listening Test (1-4 scale)}
\label{fig:similarity_test}
\end{subfigure}
\caption{Subjective listening results }
\end{figure*}

\begin{figure*}
\centering
\begin{subfigure}[b]{0.29\textwidth}
    \includegraphics[width=\textwidth]{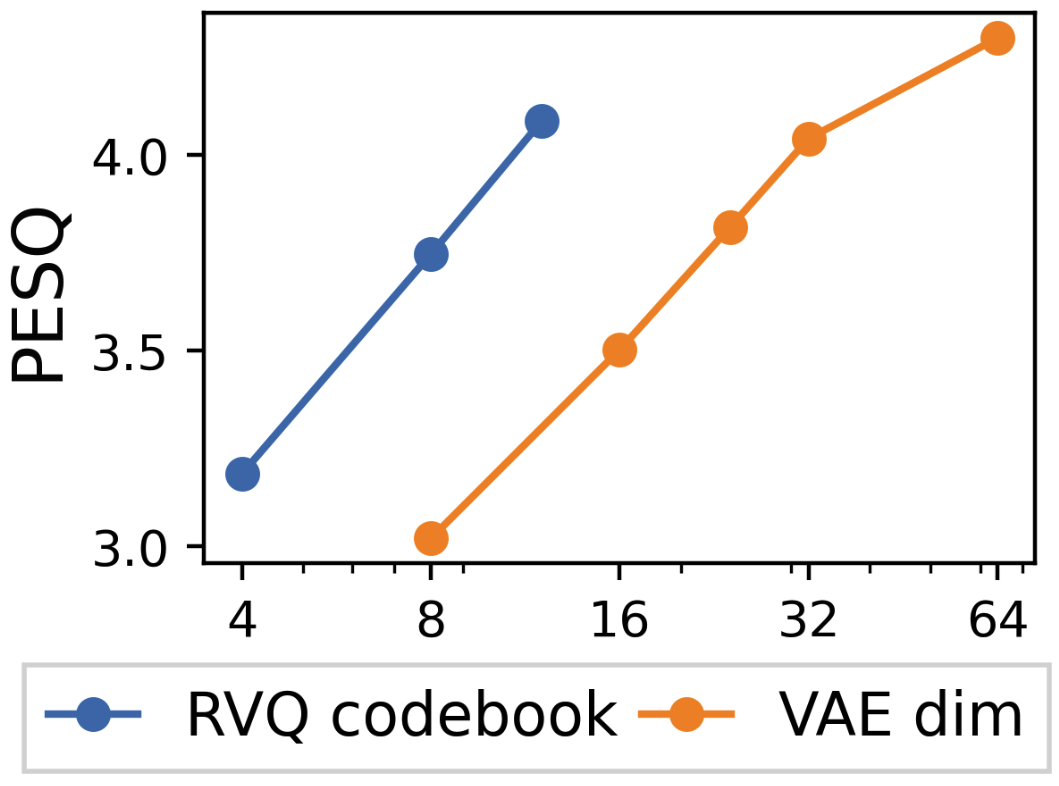}
    \caption{VQ/VAE reconstruction}
    \label{fig:pesq_reconstruction}
\end{subfigure}
\begin{subfigure}[b]{0.33\textwidth}
\includegraphics[width=\textwidth]{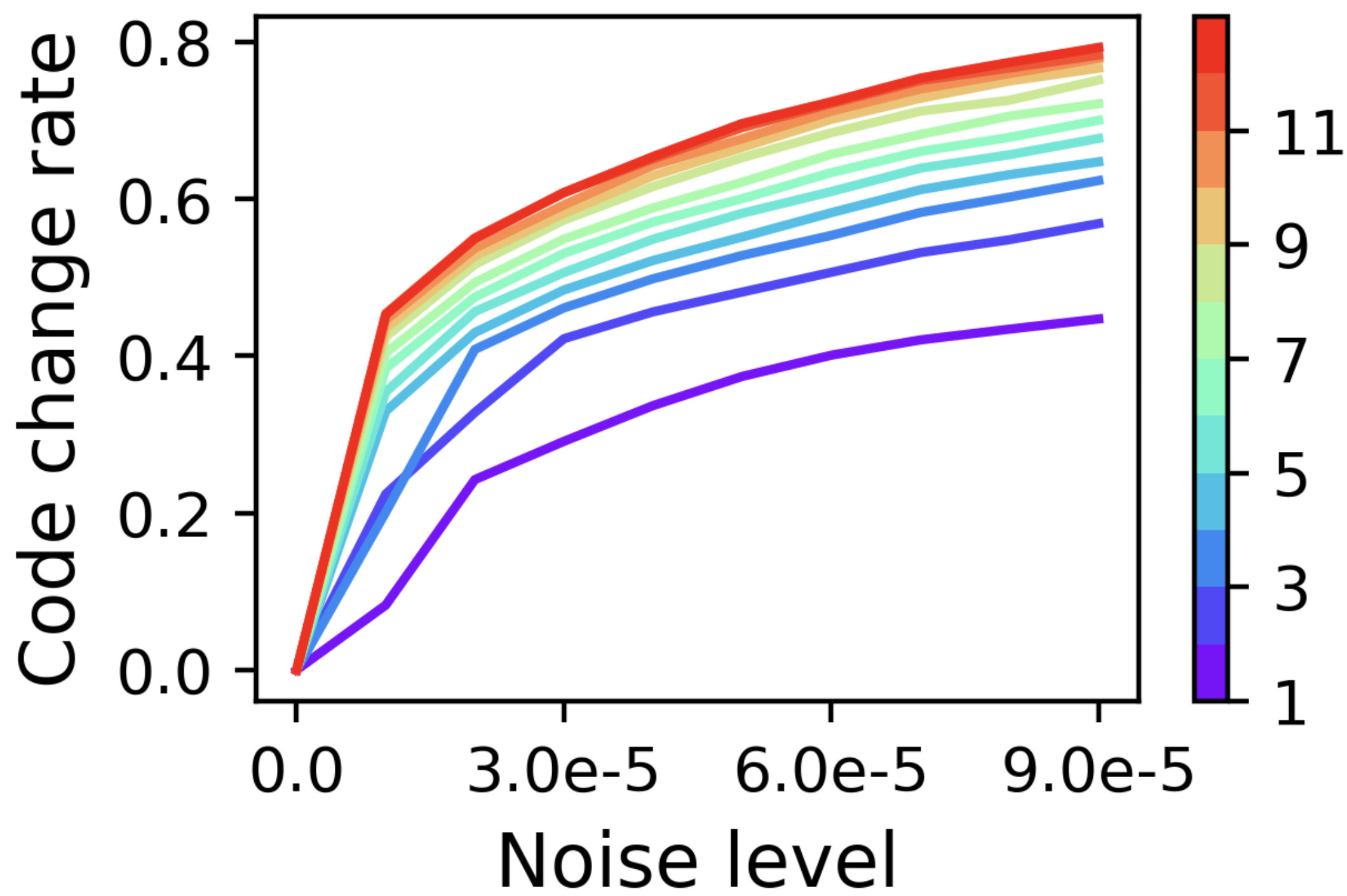}
\caption{Codewise Noise sensitivity}
\label{fig:noise_sensitivity}
\end{subfigure}
\begin{subfigure}[b]{0.29\textwidth}
\includegraphics[width=\textwidth]{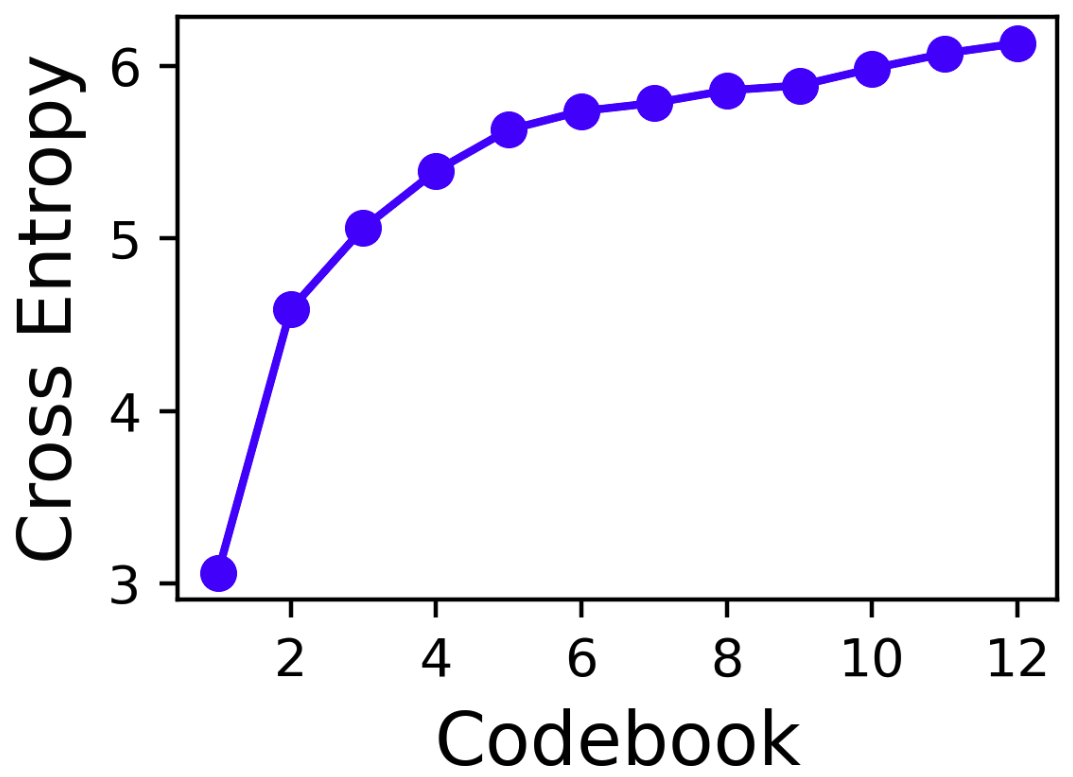}
\caption{Codewise cross entropy loss}
\label{fig:codebook_losses}
\end{subfigure}
\caption{High-fidelity RVQ codes}
\end{figure*}

\begin{table}[t!]
\setlength{\tabcolsep}{4pt} 
\centering
\small
\begin{tabular}{@{}llllll@{}}
\toprule
Model & Type & UTMOS & CER & Sim. \\ \midrule
GT & -- & 4.121 & 0.528 & 0.736 \\ \midrule
XTTS & Disc. & 3.91 & 0.787 & 0.544 \\
VoiceCraft & Disc. & 3.986 & 4.067 & 0.598 \\
\method & Disc. & 4.27 & 2.298 & \textbf{0.600} \\
\method & Cont. & \textbf{4.28} & \textbf{0.739} & 0.539 \\
\bottomrule
\end{tabular}
\caption{Objective evaluation of LibriSpeech \textit{test-clean}}
\label{table:librispeech_objective}
\end{table}

\newpara{Metrics.}\label{par:metric}
We measure \textit{Audio Quality} using UTMOS \cite{UTMOS} which produces a quality score in the range of 1-5 (higher is better). 
\textit{Intelligibility} is measured by the character error rate (CER) in percentages (\%) between the ground-truth text and the Whisper transcripts \cite{Whisper} of the synthesized audio. 
\textit{Speaker Similarity} is measured by the cosine similarity to the prompt, comparing the embedding of WavLM-TDNN \cite{wavLM}, a popular speaker verification model. This metric was also reported in Vall-E and subsequent studies \cite{VALLE, VallE2}. The similarity score predicted is in the range of $[-1,1]$, where a larger value indicates a higher similarity.

For the subjective \textit{listening tests}, we selected one random utterance for every speaker in LibriSpeech \textit{test-clean} (20 female and 20 male speakers), resulting in 40 utterances for evaluation.
For each sample, we selected a three-second-long speaker prompt from another random utterance of the same speaker. 
Each system synthesizes the desired utterance based on the same text and speaker prompt.
All experiments were conducted on the Amazon Mechanical Turk crowd-sourcing platform with votes collected from 39-58 subjects qualified as \emph{masters}~\cite{sodre2017analysis}. 

In the first listening test we assess speech quality and naturalness by the standard 5-point scale Mean Opinion Score (MOS) ~\cite{ribeiro2011crowdmos}. 
25 distinct subjects assessed each utterance.
We report the average scores and the 95\% confidence interval.

In the second listening test we asses the Speaker Similarity by a 4-level pairwise similarity test, as in~\cite{wester16_interspeech, kons2018neural}, where subjects were presented with \textit{(utterance, prompt)} pairs and asked   
to rank speaker similarity of each pair on a 4-level categorical scale \textit{(definitely different speakers, probably different speakers, probably the same speaker, definitely the same speaker)}. 
Each utterance was assessed by 20 distinct subjects on average.
We report the mean similarity score and the 95\% confidence interval while attaching 1-4 numerical values to the above categories, as in~\cite{kons2018neural}.

We further compare \method~with two leading publicly available zero-shot TTS systems: XTTS~\cite{XTTS}, run with its official package using default inference settings; and VoiceCraft~\cite{peng2024voicecraft}, using the $330M_TTSEnhanced$ model configured for top-k sampling (k = 40), nucleus sampling (p = 1), temperature = 1, and repetition penalty = 3. The VoiceCraft authors report that this configuration outperforms their reimplementation of VALL-E2\cite{VallE2}, which we do not include directly in our evaluation, as it is not publicly available and its demo set is too limited for reliable batch metrics.

\subsection{Results}
\label{sec:results}
\newpara{Subjective Evaluation.}
We conduct the two subjective listening tests, described above, to compare the following systems:
(1) Ground Truth audio  
(2) XTTSv2~\cite{XTTS}
(3) SALAD
(4) SALAD - Discrete.
Figure~\ref{fig:mos_test} reports the mean opinion score (MOS) results, suggesting that the difference between the ground-truth audio (GT) to SALAD is statistically insignificant ($p>0.01$).
Figure~\ref{fig:similarity_test} presents the speaker similarity average score with 95\% confidence intervals, suggesting similar or better speaker similarity scores for all the systems but \textit{XTTSv2}. More precise analysis with two-sided Wilkinson rank-sum test~\cite{wilcoxon1945individual}  reveals that \method~ do not differ ($p>>0.01$) from the GT in terms of speaker similarity, while SALAD-discrete is marginally better than the GT ($p=0.0105$).

\newpara{Objective Evaluation.}
\label{sec:libri_objective}
We evaluate all models on zero-shot TTS. 
Given a text and a three-second speaker prompt, which is taken randomly from another utterance of the same speaker, the model attempts to synthesize the audio with the identity and prosody similar to the prompt. 
All models use the same random prompt for each sample.

Table~\ref{table:librispeech_objective} shows that
\method~achieves superior intelligibility (lowest CER) making it the most reliable model when having to synthesize an exact text, with comparable UTMOS quality to the discrete baseline. However, in terms of speaker similarity \method~ is comparable to XTTS but the discrete baseline is superior.

\begin{figure*}
\centering
\begin{subfigure}[b]{0.31\textwidth}
    \includegraphics[width=\textwidth]{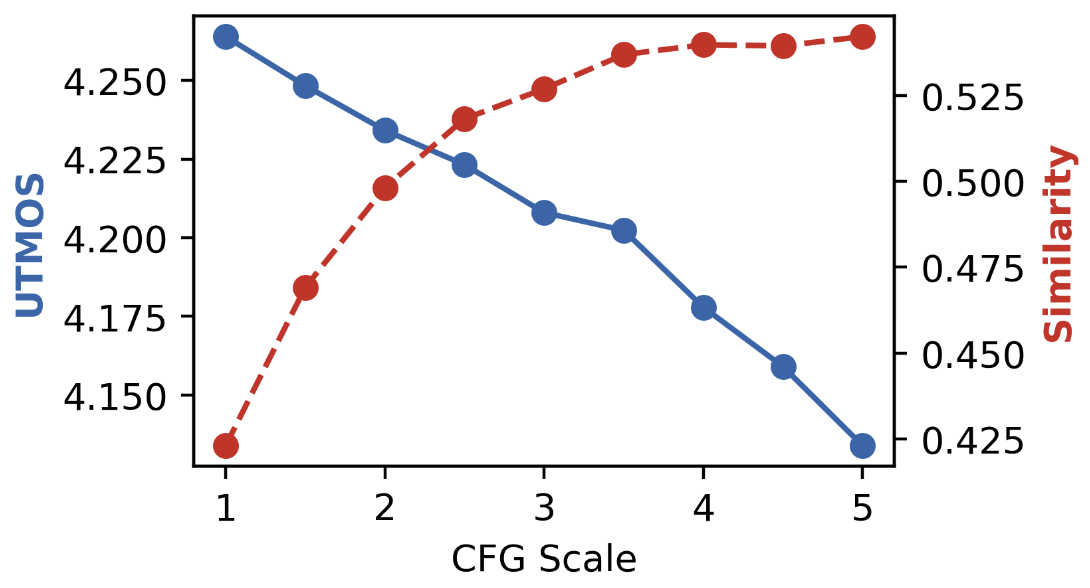}
    \caption{CFG scale}
    \label{fig:cfg_grid}
\end{subfigure}
\hspace{0.01\textwidth}
\begin{subfigure}[b]{0.31\textwidth}
    \includegraphics[width=\textwidth]{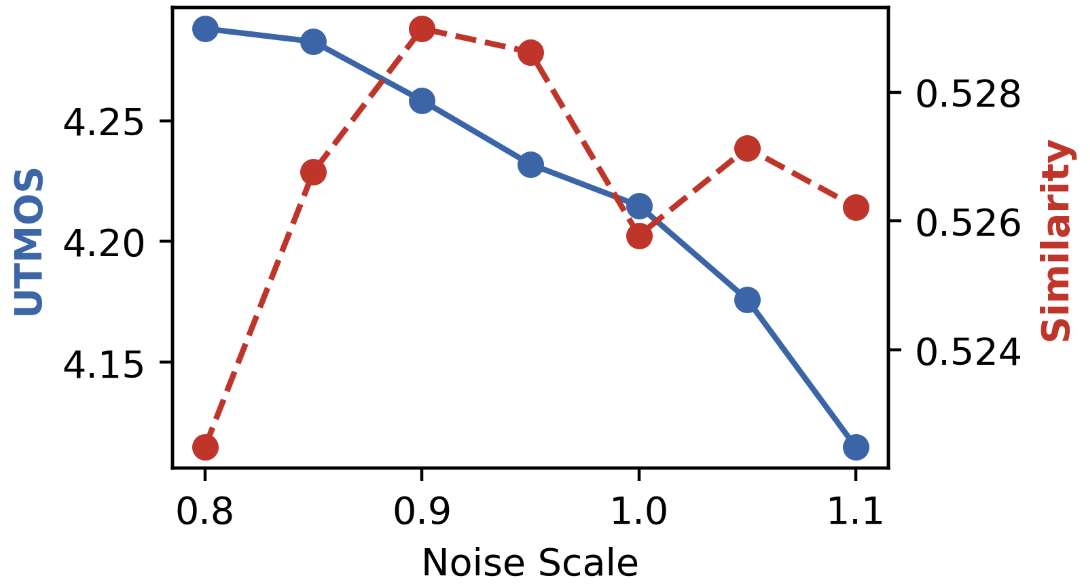}
    \caption{Noise scale}
    \label{fig:noise_scale_grid}
\end{subfigure}
\hspace{0.01\textwidth}
\begin{subfigure}[b]{0.31\textwidth}
    \includegraphics[width=\textwidth]{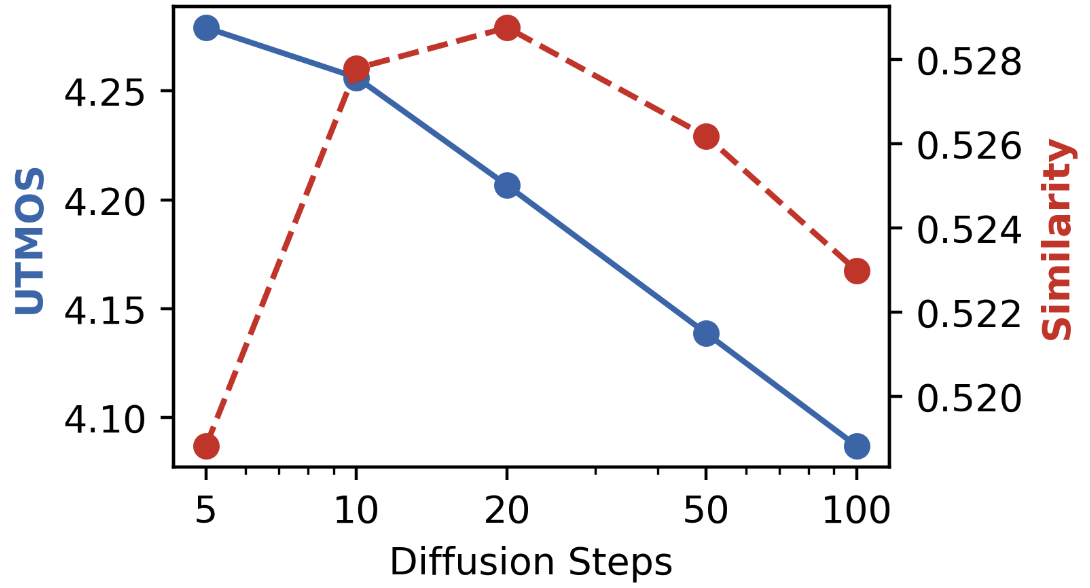}
    \caption{Diffusion steps}
    \label{fig:diffusion_steps_grid}
\end{subfigure}
\caption{Inference hyperparameters influence}
\end{figure*}

\subsection{Ablation Study}
\label{sec:ablation}
\newpara{High-Fidelity Modeling.}
\label{sec:codebook_predictability}
When increasing the number of RVQ codebooks or the VAE embedding dimension, the reconstruction quality increases, but language modeling can be difficult~\cite{NaturalSpeech2}. 
Figure~\ref{fig:pesq_reconstruction} shows the reconstruction quality measured by PESQ~\cite{PESQ}, for different numbers of RVQ codebooks and VAE embedding dimensions.
One concern regarding RVQ modeling is that the fine codes quantize noise, leading to a high gradient contribution of random classification problems.
We measure the noise sensitivity per codebook by adding Gaussian noise into raw samples, compressing them with the RVQ model, and checking the ratio of change per codebook. 
Figure~\ref{fig:noise_sensitivity} suggests that fine codebooks are more noise-sensitive.
Figure~\ref{fig:codebook_losses} shows per-codebook validation cross-entropy loss in the discrete 12-codebook model, indicating difficulty in reducing uncertainty in finer codebooks.
This occurs despite the delay pattern in Section~\ref{sec:method}, where finer RVQ levels are conditioned on coarser layers of the same frame.
Finally, we compare the generation quality with less-compressed representations. 
Table~\ref{table:high_fidelity_results} shows that higher fidelity causes a greater intelligibility drop in the high-fidelity discrete model than in the continuous model.
\

\newpara{VAE Sampling}
Unlike discrete codebooks or Mel spectrograms, VAE models enable diverse sampling, which may enhance robustness and mitigate the mismatch between training and inference(where prediction rely on previous noisy outputs). To validate this, we compare two SALAD models - one sampling from the VAE distribution $x=\mu + \epsilon\cdot\sigma$ and the other using only the mean $x=\mu$.
The results presented in Table~\ref{table:vae_sampling_influence}, show a large gap in UTMOS and intelligibility indicating that sampling improves synthesized quality. 
Listening to VAE-NoSample outputs we observed increasing speaker-inconsistency artifacts, likely due to the absence of sampling noise during training. We suspect this noise improves robustness, and we hypothesize that the high similarity score of VAE-NoSample results from these artifacts.

\newpara{Inference Hyperparameters}
\label{sec:infer_hyper_abl}
We investigate CFG, noise scale and the number diffusion steps. In every experiment, we fix all values to the default inference values following those described in Section~\ref{sec:exp_setup}, and change only a single hyperparameter.
The CFG linear extrapolation coefficient increases the speaker similarity, but degrades the automatic quality metric, as seen in Figure~\ref{fig:cfg_grid}. 
Next, we scale the noise level added in each diffusion step by scaling the $\beta_t \epsilon$ term in Equation~\ref{eq:diffusion_inference}, and see improvements in similarity but degradation in the UTMOS quality score (Figure~\ref{fig:noise_scale_grid}).
We also examine the number of diffusion steps, which improve similarity until reaching 20 diffusion steps, and also degrade UTMOS (Figure~\ref{fig:diffusion_steps_grid}).

\begin{table}
\setlength{\tabcolsep}{4pt} 
\centering
\small
\begin{tabular}{@{}llll@{}}
\toprule
 & UTMOS $\uparrow$ & Intelligibility $\downarrow$ & Similarity $\uparrow$ \\ \midrule
Continuous (d=32)     & \textbf{4.271} & \textbf{1.157}           &  0.545     \\ \midrule
Discrete (Q=12) & 4.203 & 5.461           &     \textbf{0.597}  \\
\bottomrule
\end{tabular}
\caption{Discrete vs.\ continuous \method~ with high-fidelity representations.
Intelligibility (CER) and Similarity (speaker similarity) are defined in Section.~\ref{sec:exp_setup} (Metrics).}

\label{table:high_fidelity_results}
\end{table}
\section{Discussion}\label{sec:discussion}
Compressing complex signals like audio and images involves a tradeoff between \textit{perception} and \textit{generation}. While compression can degrade perception by losing information, it benefits generation by simplifying the learned distribution. Developing generative methods for less-compressed representations could help mitigate this tradeoff. Though RVQ enables high-fidelity compression, it may quantize noise. Using continuous representations can be more robust to noise, as continuous models scale the noise according to its magnitude.

\newpara{Future work.}
Our approach could be extended by developing multimodal models capable of both perception and generation. Additionally, exploring generation stopping conditions that do not rely on discrete representations would be valuable. Adaptive inference strategies, such as adjusting the number of diffusion steps per token (e.g., using more steps for early tokens), could improve efficiency. Finally, developing quality metrics for the diffusion process could enable the use of decoding algorithms like beam search during inference.

\newpara{Limitations}
The diffusion head inference process is slower than the RVQ prediction heads, as it requires an iterative process for the token sampling.
Moreover, it does not allow to measure likelihood or confidence, which can be useful for decoding algorithms such as beam search or confidence based unmasking.
Optimal balancing of the discrete and continuous losses in \method is not easy to obtain. During training, the gradients of the discrete semantic loss increase, while the gradients of the continuous diffusion loss decrease. 

\newpara{Ethical Statement}
\method~is capable of zero-shot voice cloning, which presents potential risks, including misuse for voice spoofing. To mitigate these risks, it is essential to implement strict protocols that verify and ensure the speaker's consent and authorization before their voice is used in any real-world application, especially when dealing with previously unseen speakers. Safeguards must be in place to prevent unauthorized use and protect individuals' privacy and identity.

\begin{table}
\setlength{\tabcolsep}{4pt} 
\centering
\small
\begin{tabular}{@{}llll@{}}
\toprule
 & UTMOS $\uparrow$ & Intelligibility $\downarrow$ & Similarity $\uparrow$ \\ \midrule
Sample    & \textbf{4.280} & \textbf{0.739}           & 0.539      \\
NoSample  & 3.468 & 1.891           & \textbf{0.613}      \\ \bottomrule

\end{tabular}

\caption{Influence of VAE sampling during training}
\label{table:vae_sampling_influence}
{-0.2cm}
\end{table}

\newpara{AI Tools Usage}
AI-based tools may have assisted in code writing and paraphrasing, but all content was thoroughly reviewed by the authors and used only as a support tool.

\section*{Acknowledgment} 
This research work was partially supported by Mobileye Academic Grant Program and by the Israel Innovation Authority, grant number 78563.
\printbibliography
\end{document}